\documentclass[12pt,preprint]{aastex}

\newcommand{\Msunyr}{\mbox{M$_\odot$ yr$^{-1}$}}

\def\arcmin{$^\prime$}

\def\fasec {{\rlap.}^{\prime \prime}\hskip0.05em}

\begin{document}

\title{A Study of Compact Radio Sources in Nearby Face-on Spiral
Galaxies. II. Multiwavelength Analyses of Sources in M51}

\author{L.A. Maddox\altaffilmark{1} and J.J. Cowan}
\affil{The Homer L. Dodge Department of Physics and Astronomy, The University of Oklahoma,
440 W. Brooks St., Norman, OK  73019}
\altaffiltext{1}{Current Address: Department of Astronomy, University of Illinois at Urbana-Champaign,
103 Astronomy Bldg., MC-221, 1002 W.~Green St., Urbana, IL 61801}

\author{R.E. Kilgard}
\affil{Harvard-Smithsonian Center for Astrophysics, 60 Garden St., Cambridge, MA  02138}

\author{E. Schinnerer}
\affil{Max-Planck-Institut f\"ur Astronomie, K\"onigstuhl 17, 69117 Heidelberg, Germany}

\author{C.J. Stockdale}
\affil{Department of Physics, Marquette University, P.O. Box
1881, Milwaukee, WI  53201}

\begin{abstract}

We report the analysis of deep radio observations of the interacting
galaxy system M51 from the Very Large Array, with the goal of understanding
the nature of the population of compact radio sources in nearby spiral galaxies.
We detect 107 compact radio sources, $64\%$ of which
have optical counterparts in a deep H$\alpha$ Hubble Space Telescope image. 
Thirteen of the radio sources have X-ray counterparts from a {\em Chandra}
observation of M51.  
We find that six of the associated H$\alpha$ sources are young supernova remnants
with resolved shells.  Most
of the SNRs exhibit steep radio continuum spectral indices consistent 
with synchrotron emission.
We detect emission from the Type Ic SN~1994I nearly a decade after explosion: 
the emission ($160\pm22~\mu$Jy beam$^{-1}$ at 20~cm, $46\pm11~\mu$Jy beam$^{-1}$
at 6cm, $\alpha=-1.02\pm0.28$) is consistent with light curve models for Type Ib/Ic supernovae.
We detect X-ray emission from the supernova, however no optical counterpart is present.
We report on the analysis of the Seyfert 2 nucleus in this galaxy, including the
evidence for bipolar outflows
from the central black hole.

\end{abstract}


\section{Introduction}

M51 (NGC 5194/5) is a nearby, grand-design spiral galaxy with its early-type
companion forming an interacting pair.  It is nearly face-on ($i\sim20^{\circ}$),
which makes it an ideal candidate for population and morphology studies at
all wavelengths. This is the second galaxy in our study of nearby spiral
galaxies, with the aim to classify and understand the populations of compact
radio sources in these galaxies.  
Results of a long-term study of M83 were presented in Part I of this series \citep{madd06}.  Results
of our radio observations of M101 and NGC~3184 and a thorough comparison of all four galaxies
will be presented in Part III of this series.

The few distance determinations for M51 range from 6.0 Mpc \citep{baron96} to
9.6 Mpc \citep{sand74}.
We have opted to use the distance of 8.4 Mpc as derived by the 
planetary nebula luminosity function of the galaxy \citep{feld97}.
Despite the interaction with M51b, the star-formation rate (SFR) in the main 
galaxy M51a is only slightly elevated \citep{calz05}.
\citet{calz05} report the total SFR as ~3.4 \Msunyr and the SFR/area
as ~0.015 \Msunyr Kpc$^{-2}$, which place M51a among the "quiescently" star-forming systems.  
The M51a/b interacting system has played host to three supernovae (SNe) in modern times. 
SN~1945A occurred in M51b \citep{kowa71}, while SN~1994I \citep{puck94} and SN~2005cs 
\citep{kloe05} happened in M51a.
We will report briefly on the radio emission of SN~1994I, the only
supernova that we have detected in this galaxy with our radio observations.
An {\sc H i} study of M51 by \citet{rots90} noted that the neutral gas emission follows
the spiral arms closely.  The large quantities of gas indicate that compact radio
emission should also be concentrated along the arms.
M51 contains a low-luminosity Seyfert 2 nucleus that exhibits evidence
of a bipolar outflow \citep[e.g.,][]{ford85}.  

In this paper we present multiwavelength analyses of the numerous compact radio sources
present in M51 using new, high-resolution observations from the
Very Large Array (VLA)\footnote{The Very Large Array of the National Radio Astronomy Observatory is a facility of the National
Science Foundation operated in cooperative agreement by Associated Universities, Inc.}.  
Consistent with our previous work on  M83, we find that our detected objects fall into three
categories.
First, thermal
{\sc H ii} regions that are star forming regions, whose radio emission is dominated by thermal 
bremsstrahlung, and which have generally flat continuum spectra.  At the distances of the sample
galaxies, we expect to detect emission only from the strongest {\sc H ii} emission regions.
Secondly, we detect SNe and supernova remnants (SNRs) that
emit radio continuum emission 
through a shock interaction with the circumstellar material (CSM) or the
interstellar medium (ISM), respectively.  SNe and SNRs are characterized by a nonthermal continuum spectrum,
which fades over timescales of months to years.  The final class of 
objects we detect are X-ray binaries (XRBs) in radio-loud or flaring states.  The
emission mechanism here is the formation of bipolar jets, which interact with the magnetic
field of the compact member, producing synchrotron emission.  Like the SN/SNR emission, XRB
emission is characterized by a steep spectrum in the radio.  These sources are also transient
in the radio, going through periods of high activity after periods of radio quiescence.
In this paper, we present details of the radio observations and analysis of M51.  We 
compare our radio observations
with archived optical Hubble Space Telescope (HST) and {\em Chandra} X-ray data to find 
counterparts in these complementary bands.
Finally, we  discuss the radio emission from the Seyfert 2 nucleus.

\section{Observations and Data Reduction}

Observational parameters for this data are presented in
Table \ref{m51obs}.
The 20 cm observation was performed on 2002 April 5 for a total of 6 hours while
the VLA was in A configuration.
The 6 cm observations were performed  in B configuration over several days  in December 2003 and 
January 2004.  The total on-source integration time for the 6 cm observations was 22.5 hours.

All of the data were processed using the Astronomical Image Processing System (AIPS)
provided by NRAO.  Flux calibration was performed using 3C 286 as the primary
calibrator.  To correct for atmospheric phase variations, 
a secondary calibrator, 
J1327$+$434 was used in all observations.  
The data were then imaged using the AIPS task IMAGR
using a Briggs robustness parameter of 0.  This value has the advantage of minimizing noise
while allowing for excellent point-source detection in resultant images.  
Due to the low flux density 
in the radio sources and the low dynamic range, self-calibration
was not required.  The data sets were deconvolved
and then restored using Gaussian restoring beams, the dimensions
of which are indicated in Table \ref{m51obs}.  The size of the Gaussian
restoring beams were determined using
the values for the ``dirty'' beam, calculated from the Fourier transform of the {\em u-v} plane
coverage.  After imaging, the task PBCOR was run on each map to correct for the response of
the primary beam.

The final 20~cm radio map achieved a sensitivity of 
22.5 $\mu$Jy beam$^{-1}$ rms with a deconvolved
beam measuring $1\fasec 50\times 1\fasec 21$. The 6~cm map reached 
11.7 $\mu$Jy beam$^{-1}$ rms with
a deconvolved beam of $1\fasec 47\times 1\fasec 13$. 
Each map used a pixel scale of 
$0\fasec3$ pixel$^{-1}$
and covers a sky area of 10.2\arcmin per side.  

\section{Data Analysis}

An initial source list was obtained using the AIPS task SAD, which searches for points in 
a radio map that are higher than a specified level.  Our search consisted of four iterations,
each looking at fainter flux levels.  The flux cutoff levels were 1.0, 0.5, 0.1, 0.05 mJy beam$^{-1}$.
A Gaussian fit was applied to each detected emission source by the program, and fits 
that fail are rejected by the algorithm.  Some extended or slightly extended sources 
were listed as multiple entries.  Sources which exhibited extended emission, larger 
than 2 beam widths across ($> 2\fasec 5$, were removed.  Our source list was compressed
to account for this. Finally a visual inspection of the maps was performed in order to 
find real emission that was rejected by the detection algorithm.  This was the case 
for ten faint sources which were detected in both wavelengths at identical positions.  
The final list of sources, including peak flux densities and
spectral indices ($S\propto\nu^{+\alpha}$) is listed in Table \ref{m51flux}.
The final 20 cm radio map is shown in Figure \ref{m51maps}. The spiral arms are evident
in the radio maps, as is the position of M51b, the interacting companion galaxy.  As a 
comparison, Figure \ref{m51op} shows a three-color HST image
of M51.  

The peak flux density and rms values on the final source list were determined using the AIPS task IMFIT.  The input model was a simple two-dimensional
Gaussian.  Peak flux densities are listed in Table \ref{m51flux}.  Where the fit failed, we assume a 3 $\sigma$
upper limit of 65 $\mu$Jy at 20 cm and 36 $\mu$Jy at 6 cm.

\section{Discussion}

\subsection{Radio Point Sources}

In Table \ref{m51flux}, we present the results of our radio survey of the M51 system
listing the positions, radio flux densities at 20 cm and 6 cm (when 
detected at the appropriate band), and the spectral index ($\alpha$, 
$S_{\nu} \propto \nu^{+\alpha}$) derived from these measurements for our
107 radio point sources.  We have determined that
33 of these sources have steep indices ($\alpha < -0.3$), 45 sources
have flat indices ($-0.3 < \alpha < 0.3$), 3 sources have inverted
spectral indices ($0.3 < \alpha$). The final 26 sources cannot be distinguished
as having either a flat or steep spectral index due to the 
experimental uncertainties. Figure \ref{m51or} shows the sources overlayed on
an HST/ACS I Band image of M51. 

Sources 15, 20, 35, 40, and 78 have very steep spectral indices ($\alpha < -1$), 
with sources  40 and 78 having only been detected in the 20~cm observations.  
These sources may be background sources (e.g., unresolved radio galaxies),  
optically obscured SNRs, or highly variable, unresolved radio sources in M51, 
as the 20~cm and 6~cm data were obtained 21 months apart.  Source 20 is  associated
with an optical {\sc H ii} region, so it is likely this source is an optically obscured SNR.

All of the detected radio sources lie along the spiral arms of the galaxy.  The three inverted sources,
denoted as blue circles in Figure \ref{m51or} lie in dust lanes in the inner region of 
the galaxy. They could be embedded \ion{H}{2} regions experiencing free-free absorption
due to a high column density of gas.  
In order to test for free-free absorption in these compact sources, we would require 
high resolution radio observations at lower and higher frequencies to look for a 
spectral turnover at low frequency. A similar study was performed by \citet{ting04} for 
NGC 253, who found significant evidence for free-free absorption of the compact radio
emission regions.

The flat spectrum sources in Figure \ref{m51or}, indicated in yellow, tend to 
populate the outer spiral arms, away from dust-rich areas.  The steep spectrum sources are
seen mostly along the inner arms.  If these sources are radio SNRs, the position
along the inner region is not unusual, as star formation in this region would be higher.
The indeterminant sources also lie mostly in the inner region.  Some of these sources might
be absorbed SNRs, as they are positioned along the dust lanes.  High resolution, narrow-band
optical observations (e.g., [\ion{O}{3}] and [\ion{S}{2}]) would be required to further classify
these objects.

We have identified 44 radio sources coincident with previously detected optical \ion{H}{2} regions.  
Of these, 27 sources
have flat spectral indices, 8 have steep spectral indices and  8 have indices which cannot be classified
as either due to experimental uncertainty. One of the 45 sources, source 13,
has an inverted spectral index $>$1.0.  
This source is likely embedded in the disk and the 20~cm flux
is being absorbed by intervening gas \citep[e.g.,][]{ting04}.
Sources 17 and 34 with spectral indices of $\sim0.5$
are also likely embedded {\sc H ii} regions.

\subsubsection{Supernovae}

At the time of our observations (April 2002  and January 2004), M51 had been
the host to two historical 
supernovae: the Type I SN~1945A \citep{kowa71}, and the Type Ic SN1994I 
\citep{puck94}.
We were unable to detect emission from the reported position of
SN~1945A, which exploded in M51b and would be near the edge of our
field of view. Figure \ref{m51op} shows the position of the supernovae on
an optical DSS image of M51.

We have detected non-thermal radio emission from the reported optical
position of SN~1994I \citep[Source 58,][]{puck94}, nearly a decade after its
initial discovery 
(see Figure \ref{m5194i}).
The flux level at this time (160 $\mu$Jy at 20 cm) is consistent with a Type
Ic supernova,
based on the model of \citet{weil02}.
If we assume a uniform, progenitor wind velocity of 10 km s$^{-1}$,
we determine a  progenitor mass-loss rate of
$\dot{M}=2.75\times10^{-6}~M_{\odot}~yr^{-1}$ \citep{stock05a}.
An estimate of the progenitor mass-loss based on the X-ray emission (using
the same wind velocity)
yields a rate of $\dot{M}\sim10^{-5}~M_{\odot}~yr^{-1}$, with a reported
value for the luminosity of $\sim 10^{37}
{\rm erg s^{-1}}$ \citep{imml02}.  Our measured radio spectral index for
SN~1994I of $-1.04$ is typical for type Ic SNe and equivalent with the value
reported by \citet{weil02} for the 
early radio observations
of $-1.16$.   Ten years after explosion, the 6 cm flux ($46\pm 11{\rm \mu
Jy}$)
is beginning to fade below detectable limits of a typical deep VLA search.
 
While our observations pre-date the explosion of the type II SN~2005cs,
subsequent radio observations of this SN did not detect emission from
this source at 1.3, 2.0, 3.5, and 6.2 cm \citep{sto05}.  We detect no
identifiable radio emission from the immediate nearby region, which would
indicate the presence of radio SNRs or HII regions.

\subsubsection{Optical Counterparts}

We overlayed our radio positions on an optical HST/ACS image available 
through the Hubble Heritage project (Figure \ref{m51or}).  Of our 107 compact
radio sources, 44 are coincident with large {\scshape H ii} regions detected
in H$\alpha$.
An additional 24 have more compact H$\alpha$ counterparts associated with stellar clusters.
Six of these compact sources (5, 47, 68, 74, 76, and 84) are coincident with resolved H$\alpha$
shells (Figure \ref{regbu}). Other broad optical bands show only emission
from stars at these positions.  Two of the sources, 47 and 84, have associated X-ray emission, and
will be discussed in Section \ref{m51xr}.  The remaining sources, with the
exception of Source 5 exhibit non-thermal radio emission consistent with that
of SNRs.  The angular diameter of most remnants, as seen in the H$\alpha$ images,
 are $\sim0.3-0.5^{\prime\prime}$, corresponding to 12.2-20.4 pc.
If we assume a Cas A like expansion velocity of 6000 km s$^{-1}$, this would
give an age of $\sim 2000-3300$ yr for the largest shell source.

The remaining {\sc H ii} counterparts lie in large clouds of ionized gas.  In
other optical filters, these clouds contain  large star clusters (Figure \ref{m51clust}).
Most of these radio sources have flat spectral emission, but a few exhibit
non-thermal spectra similar to the emission from the shell sources.
Four of the sources in this category have X-ray counterparts, and will
be discussed in Section \ref{m51xr}.

To estimate the strength of the {\scshape H ii} regions in
M51, we have computed excitation parameters based on the 6 cm radio
properties that we have determined from our observations.  Using the
formulae from \citet{schra69} and \citet{mezg67}, we calculate the excitation parameter $U$:
\begin{equation}
U=4.5526\left[a(\nu,T)^{-1}\nu^{0.1}T_e^{0.35}S_{\nu}D^2\right]^{1/3},
\end{equation}
where
\begin{equation}
a(\nu,T)=0.366\left(\frac{\nu}{\mathrm{GHz}}\right)^{0.1}\left(\frac{T_e}{\mathrm{K}}\right)^{-0.15}
\left\{\ln\left[4.995\times10^{-2}\left(\frac{\nu}{\mathrm{GHz}}\right)^{-1}\right]+1.5\ln
\left(\frac{T_e}{\mathrm{K}}\right)\right\}
\end{equation}
where $\alpha$ is a unitless scaling factor of order unity.
Here, $\nu$ is the frequency of observed radiation in GHz, $T_e$ is the electron gas temperature in K,
$S_{\nu}$ is observed radio flux density in Jy, and $D$ is the distance to the cloud from the 
observer measured in kpc.
We assume an average {\sc H ii} temperature of $T_e=10^4$ K.
The results of these calculations are listed in Table \ref{exparm}.

Similarly to our findings in M83 \citep{madd06}, we are detecting only the brightest
{\sc H ii} regions in our observations.  The larger sample of optically fainter {\sc H ii} 
regions  are below are the detection threshold of our radio observations.
Eleven of the sources exhibit steep spectral indices ($\alpha < -0.3$), indicating a non-thermal
emission mechanism. Five of the source are confirmed SNRs (see Section \ref{m51xr}), and the 
remaining ones
are  SNR candidates.  High resolution {\sc [O iii]} and {\sc [S ii]} observations
are needed to confirm the classification of these sources as SNRs.

\subsubsection{\label{m51xr} X-ray Counterparts}

Many radio sources also coincide with bright X-ray sources in  
external galaxies.  Though most of the luminous X-ray sources  
detected in galaxies are X-ray binaries, few are detectable in the  
radio, assuming isotropic emission.  Radio emission from these sources
is typically very weak in the long periods of quiescence. 
It may, however, be possible to  
detect beamed emission, e.g., from a microquasar.  The presence of such an 
X-ray counterpart can aid in the classification of these radio sources.

Of the 107 radio sources detected in M51, 13 have X-ray counterparts  
from Chandra observations \citep{kil05}.  The Chandra  
observations reach a uniform limiting luminosity of less than $5 
\times\ 10^{36} erg\ s^{-1}$.  Basic properties of these 13 sources  
are listed in Table \ref{m51xro}.  As described in \citet{prest03} and  
\citet{kil05}, it is possible to assign a rough source  
classification to an X-ray source based upon its X-ray luminosity,  
variability, and placement on an X-ray color-color diagram.  Figure  
\ref{modcc} shows a model X-ray color-color diagram.  SNRs, which typically  
have thermal X-ray spectra with temperatures of a few hundred eV, are  
located in the lower-left portion of the color-color diagram.  As the  
spectrum becomes more absorbed, the source  moves up in the diagram.  The right  
half of the diagram is occupied by sources with power-law or 
multi-component spectra.  Thus, all SNRs should be found on the left side  
of the diagram, regardless of the absorbing column densities.

Figure \ref{modcc} shows the X-ray color-color diagram for the radio sources  
with X-ray counterparts.  As can clearly be seen, most of the sources  
have X-ray colors indicative of thermal spectra with increasing absorption.   
Indeed, 8 of the 13 sources are near textbook examples of SNR-like X-ray
spectra.  These sources are numbers 47, 49, 52, 58, 68, 73, 79, and 84.

Of the 8 sources with SNR-like X-ray colors, 6 are coincident with H$ 
\alpha$ shells or discrete H$\alpha$ point sources in observations  
with the Advanced Camera for Surveys (ACS) on HST (optical/X-ray  
coincidences are discussed in \citet{kil06}).  The remaining 2,  
sources 73 and 79, are associated with massive star clusters with  
embedded H$\alpha$ emission, thus determining the position of any individual SNRs in  
the H$\alpha$ is not feasible.  These sources are located along the  
spiral arms of the galaxy.  
Source 58, along the inner spiral arm, is in a confused H$\alpha$  
emission region, though we have already determined it to be emission  
from SN~1994I.

A brief discussion of each of the remaining 5 radio sources with X-ray 
counterparts, which do not exhibit SNR-like X-ray properties is provided here.

\begin{itemize}
\item {Source 4 lies outside the main disc of the galaxy.  Its X-ray colors  
are consistent with an XRB-like spectrum, though a soft AGN spectrum  
cannot be ruled out.  The source is coincident with a blue point  
source in the optical ACS images.}

\item{Source 12 has a highly absorbed X-ray spectrum, with no photons  
detected below 2 keV in any of the Chandra observations.  (It is  
located at the far right of the color-color diagram).  It also  
exhibits variability on the timescale of months to years between  
Chandra observations.  The source is coincident with a very faint red  
point source in the optical images from ACS.  The steep radio  
spectrum points to a highly non-thermal emission mechanism. All these  
things combined lead to the conclusion that the source is likely a  
background AGN whose soft X-ray emission is absorbed by M51.}

\item{Source 65 is coincident with a compact H$\alpha$ source, but exhibits  
XRB-like X-ray colors.}

\item{Source 95 is highly absorbed in the X-rays.  (It is located in the  
``Absorbed sources'' ellipse of the color-color diagram).  It is thus  
impossible to decide  between XRB or SNR spectra.  However,  
the position towards the left on the color-color diagram means that  
we can rule out hard X-ray spectra.  The source is thus consistent  
with either having a SNR-like or soft XRB-like spectrum.  There is no  
counterpart detected in B, V, I or H$\alpha$.}

\item{Source 107 is the brightest of the X-ray/radio overlaps.  It exhibits  
a flat radio spectrum between 20cm and 6cm.  This would normally  
indicate a thermal process that would also produce  H$\alpha$ emission; however, no H$ 
\alpha$ emission is detected in the ACS images.  The radio  
observations were separated by $\sim20$ months.  It is possible that  
we are observing a microquasar in a radio-loud state \citep{nipo05}.}
\end{itemize}

\subsection{Nuclear Emission}

The X-ray and radio morphologies of the nuclear region
are strikingly similar.  They consist of an area of ring-like 
emission to the north, postulated to be a bubble blown
out due to an outflow \citep{ford85}, and a dense area
of emission to the south of the nucleus.  Following the
studies by \citet{tera03} (and references therein), we
refer to the southern source as the extranuclear cloud (XNC).
The nucleus itself is unresolved.
Figure \ref{m51nuc} shows radio contours overlayed onto a
three color X-ray image of this nuclear region.

X-ray spectral modeling of the nucleus determined that the emission
consisted of two major components.  The hard component arises from reflection
of the radiation from a compact nuclear source with a power law spectrum by
cold matter in the vicinity. The soft component is similar to the XNC, being
well modeled by a thermal plasma that is shock-heated by mass outflow \citep{tera03}. 
The shock heating process results in non-thermal continuum radio emission  as
shown in \citet{crane92}.  For the nucleus (Source 52), we
measure a steep spectral index ($-0.82\pm0.27$).

We measure a similar steep spectrum for the XNC ($\alpha=-0.81\pm0.02$).
In high resolution 6 cm observations of the nucleus, \citet{crane92} identify
a radio jet emanating from the nucleus and terminating in the XNC. The jet 
drives the heating of the cloud and fuels the expansion of the XNC, as suggested by
 \citep[e.g.,][]{ford85}.  
The jet is the site of the hardest X-ray emission in the XNC
\citep{tera03} and the broadest optical line emission \citep{cecil88}.
\citet{gopal00} presented a model where shells in spiral galaxies, like those seen in our data, may be 
produced through shock heating due to a radio jet.  The interaction of the jet with the material within
the disk would lead to a bow shock, which is seen in the XNC by \citet{crane92}.
Our radio observations lack the resolution of \citet{crane92} 
and are unable to resolve the radio jet.

Comparison of the bubble in the two radio bands yield a non-thermal
spectral index measurement ($\alpha=-0.58$), which is consistent
with a synchrotron emission mechanism. This matches the X-ray analysis
 of the bubble in \citet{tera03}.
The knots of X-ray emission (not seen in the our radio observations
due to resolution limits) may be SNRs.  These sources are not heavily
absorbed, so they could also be clumps of hot gas on the edge of the 
bubble or soft X-ray binaries that are not embedded
in the diffuse emitting gas. 
Our observations and those of \citet{crane92} indicate no direct evidence for
jet emission in the bubble region of the nucleus.   The ring-like structure of the bubble
indicates cooler gas within, which along with a shallower spectral index make it
unlikely that the bubble is powered by a continuous jet.  \citet{rudn82} suggested a one-sided ejection mechanism to
explain the asymmetrical nature of some double lobed radio sources.  If this is the case, the bubble
could be a relic of a previous one-sided ejection cycle by the AGN of M51a.

\section{Conclusions} 

We have presented a multiwavelength study of the compact radio sources
in M51.  Primarily using radio, optical and X-ray observations, we have classified
most of the detected radio sources.  
Our primary findings are:

\noindent$\bullet$
We detected 107 compact radio sources, 44 with large  {\scshape H ii} counterparts, and 24 additional
 sources are associated with stellar clusters.

\noindent$\bullet$
Thirteen radio sources have X-ray counterparts.  Eight of the 13 sources have SNR-like
X-ray spectra.  Two of these X-ray/radio sources are coincident with resolved H$\alpha$
shells.  Two of the remaining sources are highly absorbed.  Source 12 is likely a background
AGN, while Source 95, though there is no optical counterpart, could be a dust-embedded 
SNR or an XRB.

\noindent$\bullet$
Six of the radio sources associated with H$\alpha$ sources are young SNRs with resolved shells.  Most
exhibit steep continuum radio spectra consistent with synchrotron emission.
Based on a Cas A expansion velocity, we estimate the age of the SNRs to be
$\sim2000-3300$ yr.

\noindent$\bullet$ 
We detected emission from the Type Ic SN~1994I nearly a decade after explosion.  We found
the emission at this epoch to be consistent with light curve models for Type Ib/Ic supernovae.
The SN was also detected in X-ray, though no optical counterpart was seen.

\noindent$\bullet$
The result of our analysis of the nuclear region, including the XNC, bubble and nucleus, are
consistent with previous studies that suggest a bipolar radio jet from
the low luminosity Seyfert 2 nucleus.  The nucleus and XNC have identical continuum spectra,
indicative of shock heated gas interacting with the jet.

This work employed extensive use of the NASA Extragalactic Database (NED).
This work was supported by NSF Grant AST-03-07279 (JJC). CJS is a
Cottrell Scholar of Research Corporation and work on this
project has been supported by the NASA Wisconsin Space Grant Consortium.
We wish to thank an anonymous referee for their very useful comments which
have improved the quality of this manuscript.
%{\bf (NEED more Grant Numbers...)}

%\bibliography{ref,refs}

\clearpage

%%%%%%%%%  Tables %%%%%%%%%%%%%

\begin{deluxetable}{lccccc}
\tablewidth{0pt}
%\tablecaption[M51 Observations]{\label{m51obs} M51 Observations}
\tablecaption{\label{m51obs} M51 Observations}
%\rotate
\tabletypesize{\small}
\tablehead{
\colhead{Observing} & \colhead{Configuration} & \colhead{Date} & \colhead{Integration} & \colhead{Clean} & \colhead{rms Noise} \\
\colhead{Band (GHz)} & \colhead{} & \colhead{} & \colhead{Time (min)} & \colhead{Beam (arcsec$^2$)} & \colhead{($\mu$Jy beam$^{-1}$)} }
\startdata
1.425          & A & 2002 Apr 5 & 359.0  & $1.50\times1.21$    &  22.5 \\
               & \multicolumn{4}{c}{\dotfill}                  &       \\
               &   & 2003 Dec 11 & 426.5 &   &  \\
               &   & 2003 Dec 12 & 212.5 &   &  \\
               &   & 2003 Dec 20 & 90.2  &   &  \\
               &   & 2003 Dec 28 & 119.0 &   &  \\
\raisebox{1.5ex}[0pt]{4.860}  & \raisebox{1.5ex}[0pt]{B} & 2003 Dec 29 & 91.5 
& \raisebox{1.5ex}[0pt]{$1.47\times1.13$}  & \raisebox{1.5ex}[0pt]{11.7} \\
               &   & 2004 Jan 2   & 120.3 &   &  \\
               &   & 2004 Jan 5   & 120.3 &   &  \\
               &   & 2004 Jan 9   & 170.5 &   &  \\
\enddata

\end{deluxetable}

\begin{deluxetable}{lccr@{$\pm$}lr@{$\pm$}lr@{$\pm$}l}
\tablewidth{0pt}

\tablecaption{
\label{m51flux}Positions, Flux Densities and Spectral 
Indices of Point Sources in M51}
\tablehead{\colhead{} & \colhead{R.A. (J2000)} & \colhead{Dec (J2000)}
 & \multicolumn{2}{c}{20cm} & \multicolumn{2}{c}{6cm} 
& \multicolumn{2}{c}{Spectral
Index}\\
\colhead{Source} & \colhead{(h m s)} & \colhead{(d m s)} & \multicolumn{2}{c}{($\mu$Jy beam$^{-1}$)}
& \multicolumn{2}{c}{($\mu$Jy beam$^{-1}$)} & \multicolumn{2}{c}{$\alpha$\tablenotemark{a}} 
}
\startdata
1    &  13  29  30.46  &  47  12  50.61  &  394  &  21  &  190  &  18  &  $-$0.59  &  0.11 \\
2    &  13  29  36.56  &  47  11  05.51  &  68   &  21  &  43   &  13  &  $-$0.37  &  0.43 \\
3    &  13  29  38.05  &  47  12  05.84  &  470  &  22  &  264  &  14  &  $-$0.47  &  0.07 \\
4    &  13  29  38.96  &  47  13  23.60  &  84   &  22  &  102  &  18  &  +0.15  &  0.32 \\
5\tablenotemark{b}    &  13  29  39.36  &  47  08  40.72  & \multicolumn{2}{c}{$\cdots$}  &  73   &  17  &  \multicolumn{2}{c}{$>+0.09$}    \\
6\tablenotemark{b}    &  13  29  43.67  &  47  10  00.95  &  56   &  21  &  32   &  13  &  $-$0.46  &  0.55 \\
7\tablenotemark{b}    &  13  29  44.05  &  47  10  22.71  &  245  &  20  &  259  &  12  &  +0.05  &  0.09 \\
8    &  13  29  45.10  &  47  13  32.25  &  164  &  21  &  93   &  13  &  $-$0.46  &  0.19 \\
9\tablenotemark{b}    &  13  29  45.17  &  47  09  56.93  &  192  &  21  &  229  &  12  &  +0.14  &  0.12 \\
10\tablenotemark{b}    &  13  29  46.40  &  47  12  33.14  &  55   &  22  &  75   &  12  &  +0.25  &  0.43 \\
11\tablenotemark{b}   &  13  29  46.74  &  47  09  40.78  &  \multicolumn{2}{c}{$\cdots$} &  61   &  13  &  \multicolumn{2}{c}{$>-0.06$}    \\
12   &  13  29  49.14  &  47  12  57.11  &  123  &  21  &  43   & 15   & $-$0.86 &  0.27 \\
13\tablenotemark{b}   &  13  29  49.43  &  47  12  40.54  &  \multicolumn{2}{c}{$\cdots$} &  217  &  11  &  \multicolumn{2}{c}{$>+0.98$}    \\
14   &  13  29  49.53  &  47  14  00.17  &  234  &  21  &  221  &  13  &  $-$0.05  &  0.11 \\
15   &  13  29  49.60  &  47  13  27.51  &  326  &  21  &  74   &  12  &  $-$1.21  &  0.17 \\
16   &  13  29  49.43  &  47 11 23.86    &  123  &  20  &  49  &  17  &  $-$0.75  &  0.38 \\
17   &  13  29  49.93  &  47  11  20.53  &  121  &  19  &  232  &  11  &  +0.53  &  0.16 \\
18   &  13  29  49.93  &  47  11  31.11  &  464  &  22  &  235  &  11  &  $-$0.55  &  0.07 \\
19\tablenotemark{b}   &  13  29  49.95  &  47  11  26.73  &  234  &  21  &  113  &  10  &  $-$0.59  &  0.13 \\
20\tablenotemark{b}   &  13  29  50.04  &  47  11  24.89  &  205  &  21  &  43  & 17   &  $-$1.20 & 0.41    \\
21   &  13  29  50.13  &  47  11  40.32  &  140  &  20  &  54   &  10  &  $-$0.78  &  0.23 \\
22\tablenotemark{b}   &  13  29  50.13  &  47  11  36.92  &  107  &  20  &  59   &  10  &  $-$0.49  &  0.25 \\
23   &  13  29  50.20  &  47  11  51.36  &  146  &  20  &  103  &  11  &  $-$0.28  &  0.17 \\
24\tablenotemark{b}   &  13  29  50.24  &  47  11  19.06  &  71   &  21  &  63   &  11  &  $-$0.10  &  0.34 \\
25   &  13  29  50.26  &  47  11  48.48  &  \multicolumn{2}{c}{$\cdots$} &  54   &  10  &  \multicolumn{2}{c}{$>-0.16$}    \\
26   &  13  29  50.30  &  47  11  22.42  &  432  &  21  &  240  &  11  &  $-$0.48  &  0.07 \\
27   &  13  29  50.32  &  47  13  57.95  &  61   &  20  &  63   &  13  &  +0.026  &  0.87 \\
28   &  13  29  50.32  &  47  11  32.91  &  111  &  17  &  63   &  8   &  $-$0.46  &  0.20 \\
29   &  13  29  50.43  &  47  11  53.26  &  \multicolumn{2}{c}{$\cdots$} &  70   &  10  &  \multicolumn{2}{c}{$>0.06$}   \\
30   &  13  29  50.45  &  47  11  45.16  &  56   &  20  &  55   &  11  &  $-$0.01  &  0.41 \\
31   &  13  29  50.45  &  47  11  27.02  &  115  &  20  &  73   &  10  &  $-$0.37  &  0.22 \\
32   &  13  29  50.46  &  47  11  37.36  &  \multicolumn{2}{c}{$\cdots$} &  59   &  9   &  \multicolumn{2}{c}{$>-0.08$}    \\
33\tablenotemark{b}   &  13  29  50.48  &  47  13  45.22  &  82   &  21  &  73   &  12  &  $-$0.09  &  0.3 \\
34   &  13  29  50.70  &  47  11  55.86  &  \multicolumn{2}{c}{$\cdots$} &  122  &  10  &  \multicolumn{2}{c}{$>+0.51$}    \\
35   &  13  29  50.95  &  47  13  43.69  &  \multicolumn{2}{c}{$\cdots$} &  84   &  12  &  \multicolumn{2}{c}{$>+0.20$}    \\
36   &  13  29  51.50  &  47  12  00.54  &  213  &  21  &  165  &  11  &  $-$0.21  &  0.12 \\
37   &  13  29  51.57  &  47  12  08.01  &  2229 &  21  &  946  &  11  &  $-$0.70  &  0.015 \\
38   &  13  29  51.68  &  47  11  57.89  &  52   &  20  &  49   &  10  &  $-$0.05  &  0.44 \\
39   &  13  29  51.73  &  47  12  01.87  &  108  &  21  &  132  &  11  &  +0.16  &  0.21 \\
40   &  13  29  51.80  &  47  11  40.43  &  385  &  20  &  \multicolumn{2}{c}{$\cdots$} &  \multicolumn{2}{c}{$<-1.99$}    \\
41   &  13  29  51.86  &  47  11  37.01  &  211  &  20  &  115  &  10  &  $-$0.49  &  0.13 \\
42\tablenotemark{b}   &  13  29  51.99  &  47  10  54.02  &  115  &  20  &  52   &  11  &  $-$0.65  &  0.27 \\
43   &  13  29  52.01  &  47  12  02.21  &  \multicolumn{2}{c}{$\cdots$} &  55   &  10  &  \multicolumn{2}{c}{$>-0.14$}    \\
44\tablenotemark{b}   &  13  29  52.01  &  47  12  42.95  &  151  &  21  &  173  &  11  &  +0.11  &  0.15 \\
45   &  13  29  52.02  &  47  12  04.42  &  \multicolumn{2}{c}{$\cdots$} &  57   &  10  &  \multicolumn{2}{c}{$>-0.11$} \\
46\tablenotemark{b}   &  13  29  52.03  &  47  12  47.20  &  64   &  20  &  62   &  11  &  $-$0.03  &  0.36 \\
47\tablenotemark{b}   &  13  29  52.08  &  47  11  26.82  &  84   &  21  &  40   &  13  & $-$0.60 &  0.41 \\
48   &  13  29  52.17  &  47  11  36.60  &  94   &  20  &  73   &  10  &  $-$0.21  &  0.25 \\
49\tablenotemark{b}   &  13  29  52.22  &  47  11  29.48  &  87   &  29  &  50   &  13  & $-$0.45 &  0.42 \\
50   &  13  29  52.35  &  47  11  36.09  &  94   &  20  &  59   &  10  &  $-$0.38  &  0.27 \\
51\tablenotemark{b}   &  13  29  52.38  &  47  12  38.51  &  \multicolumn{2}{c}{$\cdots$} &  56   &  11  &  \multicolumn{2}{c}{$>-0.13$}    \\
52\tablenotemark{b}   &  13  29  52.73  &  47  11  21.23  &  162  &  26  &  59   &  13  & $-$0.82 &  0.27 \\
53  &  13  29  52.71  &  47  11  42.73  &  2079 &  21  &  1135 &  11  &  $-$0.49  &  0.01 \\
54  &  13  29  52.79  &  47  11  39.17  &  2047 &  20  &  761  &  10  &  $-$0.81  &  0.02 \\
55\tablenotemark{b}   &  13  29  53.22  &  47  12  39.53  &  94   &  21  &  65   &  11  &  $-$0.30  &  0.28 \\
56   &  13  29  53.86  &  47  09  54.09  &  152  &  22  &  73   &  12  &  $-$0.60  &  0.22 \\
57\tablenotemark{b}   &  13  29  53.91  &  47  14  05.44  &  \multicolumn{2}{c}{$\cdots$} &  61   &  13  &  \multicolumn{2}{c}{$>-0.06$}    \\
58  &  13  29  54.12  &  47  11  30.33  &  160  &  22  &  46   &  11  &  $-$1.02  &  0.28 \\
59   &  13  29  54.24  &  47  11  32.40  &  136  &  22  &  55   &  11  &  $-$0.74  &  0.26 \\
60   &  13  29  54.24  &  47  11  23.23  &  81   &  21  &  45   &  11  &  $-$0.48  &  0.36 \\
61   &  13  29  54.32  &  47  11  29.86  &  92   &  21  &  67   &  11  &  $-$0.26  &  0.28 \\
62   &  13  29  54.72  &  47  12  36.60  &  98   &  22  &  49   &  11  &  $-$0.56  &  0.32 \\
63   &  13  29  54.84  &  47  11  59.23  &  51   &  19  &  36   &  10  &  $-$0.28  &  0.46 \\
64\tablenotemark{b}   &  13  29  54.92  &  47  11  33.00  &  133  &  20  &  72   &  10  &  $-$0.50  &  0.20 \\
65   &  13  29  54.95  &  47  09  22.41  &  406 & 21  & 280 & 17  & $-$0.30 & 0.08 \\
66\tablenotemark{b}   &  13  29  54.98  &  47  10  48.93  &  60   &  21  &  62   &  11  &  +0.03  &  0.39 \\
67\tablenotemark{b}   &  13  29  55.08  &  47  11  35.01  &  126  &  20  &  89   &  10  &  $-$0.28  &  0.19 \\
68\tablenotemark{b}   &  13  29  55.25  &  47  10  46.16  & 158 & 23  & 61 & 14  & $-$0.78 & 0.27 \\
69   &  13  29  55.31  &  47  11  38.61  &  133  &  20  &  54   &  10  &  $-$0.73  &  0.24 \\
70   &  13  29  55.33  &  47  11  36.51  &  \multicolumn{2}{c}{$\cdots$} &  59   &  10  &  \multicolumn{2}{c}{$>-0.08$}    \\
71   &  13  29  55.33  &  47  12  02.32  &  \multicolumn{2}{c}{$\cdots$} &  44   &  10  &  \multicolumn{2}{c}{$>-0.32$}    \\
72   &  13  29  55.42  &  47  11  40.67  &  \multicolumn{2}{c}{$\cdots$} &  51   &  10  &  \multicolumn{2}{c}{$>-0.20$}    \\
73\tablenotemark{b}  &  13  29  55.42  &  47  14  02.05  &  101 & 23  & 80 & 16  & $-$0.19 & 0.30 \\
74   &  13  29  55.52  &  47  12  09.92  &  98   &  20  &  54   &  10  &  $-$0.49  &  0.28 \\
75   &  13  29  55.57  &  47  13  59.82  &  175  &  15  &  78   &  12  &  $-$0.66  &  0.18 \\
76   &  13  29  55.60  &  47  12  03.08  &  80   &  20  &  37   &  10  &  $-$0.63  &  0.37 \\
77   &  13  29  55.64  &  47  11  41.57  &  \multicolumn{2}{c}{$\cdots$} &  53   &  10  &  \multicolumn{2}{c}{$>-0.17$}    \\
78   &  13  29  55.69  &  47  11  46.61  &  122  &  18  &  \multicolumn{2}{c}{$\cdots$} &  \multicolumn{2}{c}{$<-1.04$}    \\
79\tablenotemark{b}  &  13  29  55.86  &  47  11  44.48  &  319 & 25  & 230 & 13  & $-$0.26 & 0.10 \\
80   &  13  29  55.85  &  47  11  54.58  &  246  &  22  &  142  &  11  &  $-$0.45  &  0.12 \\
81   &  13  29  55.86  &  47  11  50.61  &  147  &  21  &  61   &  11  &  $-$0.72  &  0.23 \\
82\tablenotemark{b}   &  13  29  56.13  &  47  14  08.91  &  77   &  22  &  78   &  14  &  +0.01  &  0.34 \\
83   &  13  29  56.21  &  47  10  47.33  &  109  &  21  &  73   &  11  &  $-$0.33  &  0.24 \\
84  &  13  29  57.47  &  47  10  37.08  &  128 & 22  & 59 & 14  & $-$0.63 & 0.29 \\
85\tablenotemark{b}   &  13  29  58.94  &  47  14  09.00  &  98   &  22  &  89   &  13  &  $-$0.08  &  0.27 \\
86   &  13  29  59.49  &  47  11  09.96  &  \multicolumn{2}{c}{$\cdots$} &  68   &  11  &  \multicolumn{2}{c}{$>+0.03$}    \\
87\tablenotemark{b}  &  13  29  59.53  &  47  15  58.30  &  1494 &  20  &  1181 &  21  &  $-$0.19  &  0.02 \\
88   &  13  29  59.58  &  47  11  11.32  &  95   &  20  &  77   &  11  &  $-$0.17  &  0.25 \\
89\tablenotemark{b}   &  13  29  59.61  &  47  13  59.18  &  88   &  20  &  84   &  13  &  $-$0.04  &  0.27 \\
90   &  13  29  59.84  &  47  11  12.68  &  238  &  21  &  307  &  12  &  +0.21  &  0.10 \\
91\tablenotemark{b}   &  13  30  00.11  &  47  13  30.53  &  63   &  21  &  66   &  13  &  +0.04  &  0.39 \\
92\tablenotemark{b}   &  13  30  00.36  &  47  13  18.87  &  82   &  21  &  76   &  13  &  $-$0.06  &  0.31 \\
93\tablenotemark{b}   &  13  30  00.78  &  47  11  37.73  &  \multicolumn{2}{c}{$\cdots$} &  58   &  11  &  \multicolumn{2}{c}{$>-0.10$}    \\
94\tablenotemark{b}   &  13  30  00.93  &  47  09  28.90  &  61   &  18  &  73   &  13  &  +0.15  &  0.34 \\
95  &  13  30  01.27  &  47  12  43.77  &  131 & 22  & 62 & 15  & $-$0.61 & 0.29 \\
96   &  13  30  01.41  &  47  11  57.83  &  78   &  20  &  50   &  11  &  $-$0.36  &  0.34 \\
97\tablenotemark{b}   &  13  30  01.50  &  47  12  51.42  &  412  &  21  &  443  &  12  &  +0.06  &  0.06 \\
98\tablenotemark{b}   &  13  30  01.77  &  47  11  48.81  &  54   &  22  &  60   &  12  &  +0.09  &  0.45 \\
99   &  13  30  02.03  &  47  09  51.35  &  113  &  21  &  75   &  13  &  $-$0.33  &  0.25 \\
100\tablenotemark{b}  &  13  30  02.38  &  47  09  49.10  &  170  &  21  &  194  &  13  &  +0.11  &  0.14 \\
101\tablenotemark{b}  &  13  30  02.75  &  47  09  56.94  &  67   &  21  &  64   &  13  &  $-$0.04  &  0.37 \\
102\tablenotemark{b}  &  13  30  03.50  &  47  09  41.00 &   \multicolumn{2}{c}{$\cdots$} &  78   &  14  &  \multicolumn{2}{c}{$>+0.14$}    \\
103\tablenotemark{b}  &  13  30  03.95  &  47  15  33.00 &  119   &  21  &  144  &  21  &  +0.16  &  0.23 \\
104  &  13  30  05.13  &  47  10  35.78  &  9599 &  22  &  4287 &  14  &  $-$0.66  &  0.00 \\
105\tablenotemark{b}  &  13  30  07.38  &  47  13  22.30  &  144  &  21  &  175  &  15  &  +0.16  &  0.17 \\
106  &  13  30  10.85  &  47  09  40.26  &  148  &  21  &  100  &  17  &  $-$0.32  &  0.22 \\
107  &  13  30  11.03  &  47  10  40.75  & 500 & 22  & 482 & 20  & $-$0.03 & 0.06
\enddata
\tablenotetext{a}{$S_{\nu}\propto\nu^{+\alpha}$}
\tablenotetext{b}{Coincident with optical {\sc H ii} region}

\end{deluxetable}

\begin{deluxetable}{lr@{$\pm$}lc}
\tablewidth{0pt}
\tablecaption{\label{exparm} Excitation Parameters for H$\alpha$ Sources}
\tablehead{
\colhead{Source} & \multicolumn{2}{c}{Spectral Index} & \colhead{U (pc cm$^{-2}$)
}}
\startdata
5  &  \multicolumn{2}{c}{$>+0.09$}     &  242.5\\
6  &  --0.46  &  0.55  &  184.2\\
7  &  0.05  &  0.09  &  369.9\\
9  &  0.14  &  0.12  &  355.0\\
10  &  0.25  &  0.43  &  244.7\\
11  &  \multicolumn{2}{c}{$>-0.06$}     &  228.4\\
13  &  \multicolumn{2}{c}{$>+0.98$}     &  348.7\\
19  &  --0.59  &  0.13  &  280.5\\
22  &  --0.49  &  0.25  &  225.9\\
24  &  --0.1  &  0.34  &  230.9\\
33  &  --0.09  &  0.3  &  242.5\\
42  &  --0.65  &  0.27  &  216.6\\
44  &  0.11  &  0.15  &  323.3\\
46  &  --0.03  &  0.36  &  229.7\\
47  &  --0.6  &  0.41  &  198.5\\
49  &  --0.45  &  0.42  &  213.8\\
51  &  \multicolumn{2}{c}{$>-0.13$}     &  222.0\\
52  &  --0.82  &  0.27  &  225.9\\
55  &  --0.3  &  0.28  &  233.3\\
57  &  \multicolumn{2}{c}{$>-0.06$}     &  228.4\\
64  &  --0.5  &  0.2  &  241.4\\
66  &  0.03  &  0.39  &  229.7\\
67  &  --0.28  &  0.19  &  259.1\\
68  &  --0.78  &  0.27  &  228.4\\
73  &  --0.19  &  0.3  &  250.0\\
79  &  --0.26  &  0.1  &  355.5\\
82  &  0.01  &  0.34  &  247.9\\
85  &  --0.08  &  0.27  &  259.1\\
87  &  --0.19  &  0.02  &  613.2\\
89  &  --0.04  &  0.27  &  254.1\\
91  &  0.04  &  0.39  &  234.5\\
92  &  --0.06  &  0.31  &  245.8\\
93  &  \multicolumn{2}{c}{$>-0.10$}     &  224.6\\
94  &  0.15  &  0.34  &  242.5\\
97  &  0.06  &  0.06  &  442.3\\
98  &  0.09  &  0.45  &  227.2\\
100  &  0.11  &  0.14  &  335.9\\
101  &  --0.04  &  0.37  &  232.1\\
102  &  \multicolumn{2}{c}{$>+0.14$}      &  247.9\\
103  &  0.16  &  0.23  &  304.1\\
105  &  0.16  &  0.17  &  324.6\\
\enddata
\end{deluxetable}

\clearpage

\begin{deluxetable}{lr@{$\pm$}rr@{$\pm$}rcccc}
\tablewidth{0pt}
\tablecolumns{8}
%\tablecaption[X-ray Counterparts to Radio Sources]
%{\label{m51rx} X-ray Counterparts to Radio Sources}
\tablecaption
{\label{m51xro} X-ray Counterparts to Radio Sources}
\tablehead{
\colhead{} & \multicolumn{4}{c}{Flux Density} & \colhead{Spectral} & \colhead{} & \colhead{$L_x$ (0.3--2 keV)} \\
\colhead{Source} &  \multicolumn{2}{c}{20cm} & \multicolumn{2}{c}{6cm} & \colhead{Index\tablenotemark{a}} & \colhead{CXOU} & \colhead{$10^{37}$ erg s$^{-1}$} & \colhead{X-ray ID}} 

\startdata
4 & 84 & 22  & 102 & 18  & +0.15  & J132939.0+471324 & 5 & XRB/SNR \\
12 & 123 & 21  & 43 & 15  & $-$0.86  & J132949.1+471257 & 3 & ABS \\
47 & 84 & 21  & 40 & 13  & $-$0.60  & J132952.1+471127 & 6 & SNR \\
49 & 87 & 29  & 50 & 13  & $-$0.45  & J132952.2+471129 & 4 & SNR \\
52 & 162 & 26  & 59 & 13  & $-$0.82  & J132952.7+471121 & 3 & SNR \\
58\tablenotemark{a} & 162 & 26  & 93 & 14  & $-$0.45  & J132954.2+471130 & 4 & SNR \\
65 & 406 & 21  & 280 & 17  & $-$0.30  & J132955.0+470922 & 7 & XRB/SNR \\
68 & 158 & 23  & 61 & 14  & $-$0.78  & J132955.2+471046 & 1 & SNR \\
73 & 101 & 23  & 80 & 16  & $-$0.19  & J132955.4+471402 & 2 & SNR \\
79 & 319 & 25  & 230 & 13  & $-$0.26  & J132955.9+471144 & 7 & SNR \\
84 & 128 & 22  & 59 & 14  & $-$0.63  & J132957.5+471037 & 3 & SNR \\
95 & 131 & 22  & 62 & 15  & $-$0.61  & J133001.3+471244 & 2 & ABS \\
107 & 500 & 22  & 482 & 20  & $-$0.03  & J133011.0+471041 & 1 & XRB \\
\enddata

Columns are as follows: (1) Radio source number from Table 1; (2)-(3) Radio flux densities; (4) Radio spectral
index derived from flux densities; (5) {\em Chandra} source designation; (6) 0.3--2 keV X-ray luminosity
of associated X-ray source; (7) Source classification based on X-ray colors.
\tablenotetext{a}{ Coincident with SN 1994I.}
\end{deluxetable}

\clearpage

%%%%%%%%%  Figures  %%%%%%%%%%%

\begin{figure}
%\begin{center}
%\includegraphics*[width=5.0in]{fig1.jpg}
%\end{center}
\caption[Radio images of M51]{
\label{m51maps} 
20 cm VLA radio map of M51 made in the A configuration with deconvolved beam
of $1\fasec 47 \times 1\fasec 13$.  The greyscale limits are -1 to 500 $\mu$Jy.  
We note the bright double-lobed radio galaxy east southeast of the M51 system which lies 
outside the plane of the galaxy and is not discussed further.}
\end{figure}

\begin{figure}
%\begin{center}
%\includegraphics*[width=5in]{fig2.jpg}
%\end{center}
\caption{\label{m51op} Three color optical image of M51 from HST/ACS.
{\em Image Credit:} NASA, ESA, S. Beckwith, and The Hubble Heritage Team (STScI/AURA) The crosses indicate the positions of the historical supernovae in
the galaxy.}
\end{figure}

\begin{figure}
%\begin{center}
%\includegraphics*[width=5in]{fig3.jpg}
%\end{center}
\caption{\label{m51or} 
I Band image of M51 with radio sources overlayed.  We have opted to use the smoother
I band image instead of the H$\alpha$ image as a background, to illustrate the position
of the radio sources with respect to the optical galaxy.
The radio sources have been binned according to their spectral index: steep spectrum 
sources in red, flat spectrum sources in green, inverted spectrum            
sources in blue, and indeterminate flat/steep spectrum sources in yellow.}
\end{figure}

\begin{figure}
%\begin{center}
%\includegraphics*[width=4.5in]{fig4.jpg}
%\end{center}
\caption{\label{m5194i}
Radio contours of 20 cm emission overlayed on an optical HST 
image of the central
3.7 kpc M51a.  The
position of SN~1994I is marked.  VLA observations made in the B configuration, 
with a deconvolved beam of $1\fasec 47 \times 1\fasec 13$. The contour levels 
are 70, 98, 200, 280, 400, 560, 800, 1120, 1600, and 2240 
$\mu$Jy beam$^{-1}$.} 
\end{figure}

\begin{figure}
%\begin{center}
%\includegraphics*[width=2.5in]{fig5a.jpg}
%\includegraphics*[width=2.5in]{fig5b.jpg}\\
%\includegraphics*[width=2.5in]{fig5c.jpg}
%\includegraphics*[width=2.5in]{fig5d.jpg}
%\end{center}
\caption{
\label{regbu} Blowup of a small area within M51.  The H$\alpha$
image shows the positions of radio sources, indicated by circles, associated with 
optical emission.
Two of the  H$\alpha$ sources indicate shell structure indicative
of SNRs.  The remaining images show no shell emission.  The emission
in the other optical bands are predominantly due to starlight.}
\end{figure}

\begin{figure}
%\begin{center}
%\includegraphics*[width=2.5in]{fig6a.jpg}
%\includegraphics*[width=2.5in]{fig6b.jpg}\\
%\includegraphics*[width=2.5in]{fig6c.jpg}
%\includegraphics*[width=2.5in]{fig6d.jpg}
%\end{center}
\caption{
\label{m51clust} Many sources, like source 89 shown above lie in dense stellar clusters. Source
89 (circled in the  H$\alpha$ image) is  a flat spectrum radio source which is coincident with a large {\sc H ii} region.
In I, V, and B bands it is quite evident that the photoionization of the H$\alpha$ cloud is
powered by the stars in the cluster.}
\end{figure}

\begin{figure}
%\begin{center}
%\includegraphics*[width=3.0in,angle=-90]{fig7a.jpg}
%\includegraphics*[width=3.0in,angle=-90]{fig7b.jpg}
%\end{center}
\caption{
\label{modcc}{\em (left)} Model X-ray color-color diagram.  The purple lines are power-law  
spectra with photon index increasing from left to right ($\Gamma$ = 1,  
1.5, 2, and 2.5) and absorption increasing from bottom to top (0 to  
$10^{24} cm^{-2}$).  The orange lines are thermal spectra with  
temperature increasing from left to right (0.5, 1, 1.5, and 2 keV)  
and absorption increasing as with the power-law spectra.  The light  
blue lines are power-law plus blackbody spectra.  The ellipses mark  
the likely location of thermal SNRs, X-ray binaries and sources with  
very heavily absorbed spectra (where color-classification becomes  
difficult). {\em (right)} X-ray color-color diagram following \citet{kil05} for 
our radio sources with X-ray counterparts.}
\end{figure}

\begin{figure}
%\begin{center}
%\includegraphics*[width=5in]{fig8.jpg}
%\end{center}
\caption{
\label{m51nuc} An enlargement of the nuclear region of M51.  The colors represent
X-ray emission, while the contours represent 6 cm radio emission.
The smallest point-like feature in each image is $\sim1.0^{\prime\prime}$.
The radio contour levels are 70, 98, 200, 280, 400, 560 and 800 $\mu$Jy beam$^{-1}$.
}
\end{figure}

\end{document}